\documentclass[aps,pre,showpacs,amsmath,amsfonts,amssymb,superscriptaddress,preprint]{revtex4}
\usepackage[dvips]{graphicx}
\usepackage{graphics}
\usepackage[latin1]{inputenc}

\everymath{\displaystyle}
\newcommand{\beq}{\begin{equation}}
\newcommand{\eeq}{\end{equation}}

\newcommand{\bi}{\begin{itemize}}
\newcommand{\ei}{\end{itemize}}

\newcommand{\affA}{Aix-Marseille University, Marseille, France}
\newcommand{\affB}{CNRS Centre de Physique Théorique UMR7332,
13288 Marseille, France}
\newcommand{\affC}{Centre d'Immunologie de Marseille-Luminy,
13288 Marseille, France}
\newcommand{\affD}{CNRS, UMR7280, Marseille, France}
\newcommand{\affE}{INSERM, U1104, Marseille, France}

\begin{document}


\title{Experimental assessment of the contribution of electrodynamic
  interactions to long-distance recruitment of biomolecular partners:
  Theoretical basis}

\author{Jordane Preto}
\email{preto@cpt.univ-mrs.fr}
\affiliation{\affA}
\affiliation{\affB}
\author{Elena Floriani}
\email{floriani@cpt.univ-mrs.fr}
\affiliation{\affA}\affiliation{\affB}
\author{Ilaria Nardecchia}
\email{i.nardecchia@gmail.com}
\affiliation{\affA}\affiliation{\affB}\affiliation{\affC}\affiliation{\affD}\affiliation{\affE}
\author{Pierre Ferrier}
\email{ferrier@ciml.univ-mrs.fr}
\affiliation{\affA}\affiliation{\affC}\affiliation{\affD}\affiliation{\affE}
\author{Marco Pettini}
\email{pettini@cpt.univ-mrs.fr}
\affiliation{\affA}\affiliation{\affB}

\begin{abstract}

Highly specific spatiotemporal interactions between cognate molecular partners essentially sustain all biochemical transactions in the living matter. That such an exquisite level of accuracy may result from encountering forces solely driven by thermal diffusive processes is unlikely. Here we propose a yet unexplored strategy to experimentally tackle the long-standing question of a possibly active recruitment at a distance of cognate partners of biomolecular reactions via the action of resonant electrodynamic interactions. We considered two simplified models for a preliminary feasibility investigation of the devised methodology. By taking advantage of advanced experimental techniques nowadays available, we propose to measure the characteristic encounter time scales of dually-interacting biopartners and to compare them with theoretical predictions worked out both in the presence or absence of putative long-range electromagnetic forces.

\end{abstract}

\date{\today}


\maketitle

\section{Introduction}

Living matter hosts a huge number of molecular players (i.e. proteins, nucleic acids) involved in simultaneous yet specific chemical reactions, despite an apparent lack of systematic spatial order. A phenomenological description of these biomolecular machineries at work
often makes use of the concept of "recruitment", leaving usually unclear how biomolecules partners encounter or move toward their specific targets and sites of action.
On this point significative progress has been made about DNA-protein interaction at {\it short distance}. This followed the puzzling problem posed by the E.Coli {\it lac} repressor-operator protein which was found to locate its specific DNA-binding site several orders of magnitude faster than the upper limit estimated for a diffusion-controlled process \cite{riggs,barkley}. A widely accepted approach to tackle this problem is the so-called {\it facilitated diffusion}, on which a vast literature exists (see for instance Refs.\cite{hippel1,hippel2,hippel3,cherstvy} and citations therein).
To the contrary, for DNA-protein interactions and, more generally, for any dually-interacting biomolecules the mutual approach from a {\it long distance} is addresses to as $3D$ bulk diffusion and is not further studied (by "long distance" it is meant: much larger than the Debye screening length). Actually,
at first inspection, the mutual approach of cognate partners might well be driven by Brownian motion only, as at living temperature the ubiquitously distributed water molecules move chaotically in space, colliding
with larger/heavier fluid components.
On the latter, the total outcome of many simultaneous hits are forces of both random intensity and direction.
Hence, by displacing themselves in a diffusive way through  the inner cellular space, large molecules sooner or later will encounter their targets.

A complementary proposal, which remains hitherto largely unexplored, is the possibility for
molecules to interact at a distance via the electromagnetic field which is known to have sizeable
magnitudes in living matter \cite{cifra,pokornybook}. In particular, electromagnetic attractive forces acting on
a long range might, in specific conditions, facilitate the encounters of cognate partners, so that specific biomolecular reactions would occur more effectively than if dependent on stochastic motion only. Exploring this possibility,
it should be stressed that the static dielectric constant $\varepsilon_s$ of water is particularly
high, $\varepsilon_s \simeq 80$, at physiological temperatures. In addition to this dielectric screening, freely moving ions in the cellular medium tend to make the environment electrically neutral; accordingly,
the Debye length in a biological environment is found to be smaller than $\simeq 10  \mathrm{\AA}$, as  was estimated on the basis of typical ionic strength of the cytosol \cite{cherstvy,watson}.
Electrostatic interactions
between electrically charged molecules at a distance larger than the Debye length are very unlikely.
Conversely this is not necessarily the case for electrodynamic interactions \cite{craig,roberto} since the dielectric constant depends on the frequency of the electric waves under consideration. Among
the latter, the interactions occurring between oscillating electric dipoles are of a particular interest
since in many cases the long range nature of the interaction potential is essentially ``activated'' by the proximity of the dipole frequencies (resonance). In other words, two molecules whose dipole moments oscillate at the same frequency may undergo a so-called resonant interaction \cite{stephen,mcLachlan}, which is described by the potential $U(r)\propto - 1/r^3$ with $r$ the intermolecular distance (see Appendix). On the contrary, an
off-resonance situation would produce a standard van der Waals-like potential,
\textit{i.e.} $U(r)\propto - 1/r^6$, typically a short range interaction (see Appendix). Such a frequency-selective
interaction, when applied to a biological context, might be of utmost relevance during the approach
of a molecule toward its specific cognate partner(s). To the best of our knowledge this proposition dates back to Jordan who advanced the idea that resonant interactions within a quantum framework
could play a significant role in autocatalytic reactions or influence the process of biological synthesis
in such a way that replicas of molecules present in the cell are formed \cite{jordan}. His theory was questioned by Pauling \cite{pauling}, who estimated
that such forces, supposed to occur only between identical molecules, could not be
large enough to cause a specific attraction between proteins under the thermal
conditions of excitation and perturbation prevailing in living organisms.
Other attempts to explain biological selectivity have been made later on the basis
of usual van der Waals forces \cite{pauling2,Jehle}.
In parallel, in 1968, H. Fr\"ohlich proposed a dynamical model \cite{frohlich1} to account for the capacity of biological systems to self-regulate, emphasizing that, under specific conditions of energy supply to these systems,
part of this supply would not be totally thermalized but would be used to create order in response
to environmental perturbations \cite{nota1}. In particular, the normal polarization modes of
a macromolecule (or of a part of it) may undergo a condensation phenomenon, characterized
by the emerging of the mode of lowest frequency containing nearly all the energy
supply \cite{frohlich1}. Then, relying on this model, Fr\"ohlich suggested  \cite{frohlich2,frohlich3,frohlich4} that - when occurring between two biomolecules - such dipole
oscillations could be excited enough to overcome thermal noise leading to the above mentioned frequency-dependent forces.
Fr\"ohlich's seminal work has stimulated many theoretical investigations until our present days
(see for example Refs. \cite{pokornybook,rowlands,tuszynski,reimers}).
Moreover, a vast literature is available about the experimental observation of low-frequency modes in the Raman and far infrared (TeraHertz) spectra of proteins \cite{proteins} and DNA \cite{dna}. These spectral features are attributed to collective oscillation modes
of the whole molecule (protein or DNA) or of a substantial fraction of its atoms. A-priori these collective oscillations of the molecular electric dipole moment could activate the mentioned long-distance attractive and selective recruitment interactions.
However, a clear-cut experimental confirmation of the existence of the latter ones within a biological context at the molecular level is still lacking.

In the present paper we consider a yet unexplored strategy to experimentally test, at least in simplified systems,
whether these long-range recruitment forces are actually at work between typical actors of the broad variety
of biomolecular reactions in living matter. On the basis of theoretical computations resorting to elementary
and standard methods in the theory of stochastic processes on the one side, and recent progress on
experimental methods on the other side, we make a first step toward the design of experiments to test whether such forces are actually at work in living matter.

In Section \ref{reactions} we use two dynamical models to highlight qualitative and quantitative changes between
Brownian and non-Brownian encounters of the macromolecular partners of a generic biochemical reaction.
In Section \ref{applications}, we apply our models to the case of attractive electrodynamic potential $U(r)\propto - 1/r^3$ expected to have effects at long distance, and then we report the numerical results that have been obtained with realistic parameters. Finally, in Section \ref{conclusions} we discuss how our findings can be used to design an experiment and we conclude that Fluorescence Cross Correlation Spectroscopy (FCCS) is an appropriate experimental tool to perform real-time measurements of the association kinetics of dually-interacting biopartners. It thus seems experimentally feasible to answer the basic questions formulated above by comparing the outcomes of the prospected experiments versus the theoretically predicted curves at different concentrations of the reactants.

\section{First passage time models}
\label{reactions}

\subsection{Generalities}

Our idea is in principle a natural one: different kinds of forces must have different dynamical effects.
Thus we attempted to devise an experimental protocol in order to discriminate between the dynamics of purely random encounters between reaction partners versus encounters driven by both a stochastic force plus a deterministic long-range force.
Then, by experiments resorting on available techniques, we wondered whether it could be possible to discriminate between these different dynamical regimes.

A natural way to proceed from the theoretical standpoint, that may be closely related to experimental
as well as physiological conditions, is to consider an aqueous environment, initially containing
$N_A$ particles of a species $A$ and $N_B$ particles of a species $B$.
Each molecule $A$ is expected to interact with each molecule $B$ in two ways :

\begin{itemize}
\item As soon as the distance between $A$ and $B$ diminishes below a threshold $\delta$, a biochemical reaction
instantaneously takes place, so that the two molecules are not functional anymore and are
considered as out of the system.
\item The particle $A$ and the particle $B$ interact at a distance via a two-body potential $U(r)$ of an
electrodynamic type, as long as the two molecules do not get closer than the distance $\delta$.
\end{itemize}

From a general point of view, the equations describing the dynamics of the system include both random and
deterministic forces, and therefore may be given in the form:

\begin{equation} \label{mol_dyn-full}
\begin{array}{l}
 m_A  \frac{d^2 \mathbf{r}_{A,i}}{dt^2} = - \gamma_A  \frac{d \mathbf{r}_{A,i}}{dt}
- \sum_{j=1}^{N_B} \boldsymbol{\nabla}_A U \left(| \mathbf{r}_{A,i} - \mathbf{r}_{B,j}| \right) +
 \sqrt{2 \gamma_A k T} \boldsymbol{\xi}_{A,i}(t)  \vspace{0.3cm} \\
 m_B  \frac{d^2 \mathbf{r}_{B,j}}{dt^2} = - \gamma_B  \frac{d \mathbf{r}_{B,j}}{dt}
- \sum_{i=1}^{N_A} \boldsymbol{\nabla}_B U \left(| \mathbf{r}_{A,i} - \mathbf{r}_{B,j}| \right) +
 \sqrt{2 \gamma_B k T} \boldsymbol{\xi}_{B,j}(t), \vspace{0.3cm} \\
\hspace{7cm} i = 1,\dots, N_A, \  \text{and} \  j= 1,\dots, N_B.
\end{array}
\end{equation}

Here, $m_A$, $m_B$ correspond to the masses, $\mathbf{r}_{A,i}$, $\mathbf{r}_{B,j}$ to the positions, and
$\gamma_A$, $\gamma_B$ to the friction coefficients of the constituents
of each species. $T$ stands for the temperature in the solution, and $k$ is the Boltzmann constant.
$\boldsymbol{\xi}(t)$ is the random process modeling the fluctuating force due to the collisions with water molecules,
usually represented as a Gaussian white noise process for which
$\langle \xi^{\ \alpha}_{A,i}(t) \xi^{\ \beta}_{A,k}(t')\rangle =
\delta_{\alpha\beta} \delta_{ik} \delta(t-t')$, where $\alpha,\beta=1,2, 3$ are related to each component of the
$\boldsymbol{\xi}_{A,i}$'s. The same relation is valid also for the $\boldsymbol{\xi}_{B,j}$'s.

A-priori, equations (\ref{mol_dyn-full}) describe a very complex dynamics, even in the absence of randomness. For example, assuming that the potential is, for each pair of molecules, of the form
$U (r)=c_1/r^m - c_2/r^n$, with $c_1,c_2$ constants, $m,n\in{\mathbb N}$, $m>n$
and $n\le d$ (long-range condition if $d$ is the spatial dimension),
the Hamiltonian subset of this system is actually a
nonlinear classical $N$-body system whose phase space is entirely filled with chaotic trajectories \cite{Pettinibook}.
At this stage, the addition of random forces may imply that the representative point of the system  nontrivially wanders
in phase space, despite the presence of dissipative terms which, in principle, would generate trivial attractors.
Indeed, in the overdamped limit, when the acceleration terms can be neglected, one is dealing with a randomly
perturbed first-order nonlinear dynamical system which, as integrability is exceptional, is expected
to display a complex (chaotic) dynamics. Nevertheless, instead of undertaking the numerical integration of Eqs.(\ref{mol_dyn-full}), we decided to look, as a first step, for some analytic
result that can be obtained at the cost of some simplification of the system.

Because the reaction between two particles $A$ and $B$ occurs the first time they come sufficiently close
together, we will have to focus on first passage times of a simplified version of system (\ref{mol_dyn-full}).
Generally, first passage or first return time statistics are difficult to examine in dynamical systems, then leading one to
model the system under study by keeping only its salient characteristics, either in a deterministic or stochastic manner \cite{elena1,elena2}. Here we rather choose to still work with equations (1), but to reduce drastically the dimensionality of the system then leading to keep in the model under study its salient characteristics only. This was achieved by noting that Eqs. (\ref{mol_dyn-full}) describe
the mutual interaction between the two sets of particles, $A$ and $B$, but neither the $A$
nor the $B$ particles interact among themselves.
The trajectories of the $A$ particles are indirectly coupled only through the dynamics of the $B$ particles. Thus, as a first simplifying hypothesis, we assumed that the $B$ particles are fixed, and as a consequence the dynamical behaviors of the different $A$ particles are independent.

This leads to the decoupling of the individual equations in (\ref{mol_dyn-full}) and hence to the introduction of a one-dimensional model representative of the generic dynamics of a single $A$ particle. This model is considered below according to two different versions and solved according to standard methods \cite{Gardiner}.

\subsection{Model 1: absorbing plus reflecting boundaries} \label{model}

Let us consider one fixed molecule $B$ located at the position
$z=0$ and one molecule $A$, initially located at $z = x$ (see Figure {\text 1}).
We first suppose that if $A$ reaches the boundary $z = L$ of the domain, it is reflected back to $z < L$;
whereas when $A$ reaches the position $z = \delta$ for the first time, it is absorbed.
The random trajectory $z(t)$ of the molecule $A$ may be given, as previously, in the form

\begin{eqnarray}\label{mol_dyn}
\left \{
\begin{array}{l}
 {\displaystyle\frac{dz}{dt} } = v\ , \vspace{0.3cm}\\
m \,{\displaystyle \frac{dv}{dt} } =  - \gamma v  +  F \left(z\right)
+ \sqrt{2 \gamma k T} \ \xi(t).
   \end{array}
\right.
\end{eqnarray}

\bigskip
For times much larger than the characteristic time $m /\gamma$, equations (\ref{mol_dyn}) will then relax
to a state in which $dv/dt\rightarrow 0$. This approximation is justified by the fact that the biomolecules involved in reactions of interest (protein-protein or DNA-protein)
typically weigh thousands of Daltons, and thus the characteristic relaxation times in aqueous medium are very short. Therefore, equations \eqref{mol_dyn} for the $A$ particle can be simplified as

\beq\label{lang_adiab}
\frac{dz}{dt}  = \,\frac{{F}(z)}{\gamma}  + \sqrt{\frac{2 k T}{\gamma}} \, { \xi}(t)\ .
\eeq

As it is well known, the one dimensional Langevin initial value problem \cite{Gardiner}
\beq \nonumber
\frac{dz}{dt} = a(z,t) +  b(z,t) \xi(t) \ ,   \hspace{0.4cm} z(t_0) =x,
\eeq
is equivalent to the Fokker-Planck equation (FPE) for a probability $p(z,t|x,t_0)$ of finding
the particle at $z$ at time $t$, given it was at $x$ at $t_0\leq t$

\beq
\frac{\partial}{\partial t} p(z,t|x,t_0) = - \frac{\partial}{\partial z}[a (z,t)  p(z,t|x,t_0) ] + \frac{1}{2} \frac{\partial^2}{\partial z^2} \left [ b(z,t)^2   p(z,t|x,t_0)  \right ] \ .
\eeq
From \eqref{lang_adiab}, one thus obtains

\beq \label{smolu}
\frac{\partial}{\partial t} p(z,t|x,t_0) = - \frac{1}{\gamma}\frac{\partial}{\partial z} \left[ F(z) p(z,t|x,t_0) \right ]  + \frac{kT}{\gamma}\frac{\partial^2}{\partial z^2} p(z,t|x,t_0) ,
\eeq

which is also known as the Smoluchovski equation.
\bigskip

\begin{figure}[h!]
 \begin{center}
 \includegraphics[width=4.6in]{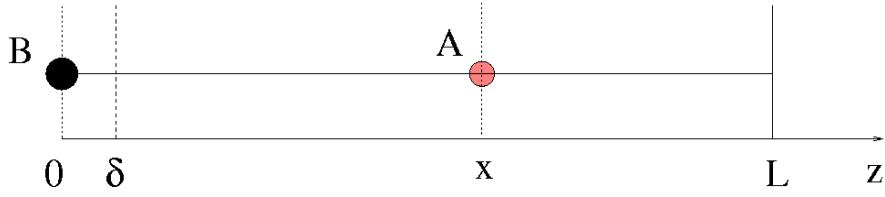}
 \end{center}$ $
\caption{\label{initAB}(Color online) A generic initial condition of Model 1 ($t=0$). Here $x$ is the initial distance
between $A$ and $B$; $\delta$ is the distance at which $A$ and $B$ react, and $L$ is the position of the reflecting
barrier for $A$ and the position of $B$ is fixed.}
\end{figure}

We now look at the time $\cal T$ at which the reaction between $A$ and $B$ occurs. That is,
the first time when particle $A$ reaches $z = \delta$.
Since we are considering an absorbing barrier at $z = \delta$ and a reflecting barrier at $z = L$,
the probability ${\mathbb P}({\cal T}\ge t)$ and the one that the particle would still be in the
interval $[\delta,L]$ at time $t$ are the same

$$
{\mathbb P}({\cal T}\ge t) = \int_{\delta}^{L}\,dz\,p(z,t|x,0) := G(x,t).
$$


Besides, as $F(z)$ and $kT$ do not explicitly depend on $t$, $p(z,t|x,0)$, and thus $G(x,t)$, are homogeneous processes, such that

\beq\label{G(x,t)}
G(x,t) = \int_{\delta}^{L}\,dz\,p(z,0|x,-t).
\eeq

This implies that $G(x,t)$ satisfies the same partial differential equation of $p(z,0|x,-t)$  for $z$ fixed, that is,
a backward Fokker-Planck equation

$$
\frac{\partial}{\partial t} p(z,0|x,-t) =   \frac{1}{\gamma}  F(x) \frac{\partial}{\partial x}  p(z,0|x,-t)  +
\frac{kT}{\gamma} \frac{\partial^2}{\partial x^2}  p(z,0|x,-t),
$$

leading to
\beq\label{eqG}
{ \frac{\partial}{\partial t} G (x,t) = - \frac{1} {\gamma} \left \{ F(x) \frac{\partial}{\partial x}   G(x,t) - kT \frac{\partial^2}{\partial x^2} G(x,t) \right \}}.
\eeq

Here, the initial condition $p(z,0|x,0) = \delta(x-z)$ (here $\delta$ is the Dirac functional) clearly gives

\beq\label{initG}
G(x,0) = 1  , \hspace{0.4cm} \text{if} \ \ \delta\le x\le L; \hspace{0.4cm} \text{and} \ \  G(x,0) = 0 ,
\hspace{0.4cm} \text{if not} ,
\eeq

whereas the absorbing condition at $\delta$ and the reflecting boundary condition at $L$ allow to write,   respectively

\beq\label{absrefG}
G(\delta, t) = 0 \hspace{0.4cm} \text{and} \ \ \left.\frac{\partial}{\partial x} G(x,t)\right\vert_{x=L} = 0 \hspace{0.4cm}, \forall t>0.
\eeq

If one focuses on the mean first passage time $\tau(x)$, which represents a characteristic time scale of
the reaction, one has by definition

\beq \label{taux}
\tau(x) =  \int \limits_0^{\infty} t \frac{\partial}{\partial t}{\mathbb P}({\cal T} < t) dt =
 - \int \limits_0^{\infty} t \frac{\partial}{\partial t} \ G(x,t)\,dt =  \int \limits_0^{\infty}  G(x,t) dt,
\eeq

after integration by parts. Then, by integrating Eq.(\ref{eqG}) between $t=0$ and $t=\infty$, and using the fact that $G(x,0) = 1$ and $G(x,\infty)=0$, we find that $\tau(x)$ must satisfy the following ordinary differential equation

$$
 - 1 =  \frac{1}{\gamma} \left \{F(x) \frac{d \tau(x)}{dx} - kT \frac{d^2 \tau(x)}{dx^2} \right \}
$$

with boundary conditions $\tau(\delta) = {\partial \tau(x)}/{\partial x}|_{x=L}  = 0$, as it follows from equations \eqref{absrefG} and \eqref{taux}. The solution is found to be \cite{Gardiner}

$$
\tau(x) = \, \int_\delta^x \,dy\,\frac{1}{\psi(y)} \int_y^L \,dz\, \frac{\gamma }{kT} \psi(z),
$$

with

\beq
\psi(x)=  \exp\left[ \int_\delta^{x} \frac{F(s)}{kT} \,ds\right]  = \exp\left\{ -\,\frac{U(x) - U(\delta)}{kT} \right\},
\eeq

\medskip

since $F(x) =  -{\partial U(x)}/{\partial x}$. This gives for $\tau(x)$:

\beq \label{Tx}
{\tau(x) = \frac{\gamma}{kT} \int_\delta^x \,dy\,\exp\left(\frac{U(y)}{kT} \right)
\int_y^L \,dz\,\exp\left( -\frac{U(z)}{kT} \right)}\ .
\eeq

\medskip

It can easily be checked that the mean first-passage time in presence of an attracting deterministic
potential, generically written as  $U(x) \propto -  x^{-n}$  with a given $n > 0$, is
smaller than the mean first-passage time with Brownian motion only, \textit{i.e.}, when
$U = 0$. Since $\exp\left( x^{-n}\right)$ is a decreasing function of $x$, we can
find an upper limit for the second integral and thus get
$$
\begin{array}{l}
{\displaystyle \tau(x) < \frac{\gamma}{kT} \int_\delta^x \,dy\,\exp\left( - \frac{1}{kT\,y^n} \right)
\int_y^L \,dz\,\exp\left( \frac{1}{kT\,y^n} \right)} \\ \\
{ \hspace{5cm} \displaystyle = \frac{\gamma}{kT} \int_\delta^x \,dy\,\int_y^L \,dz := \tau(x)^{Bwn}}.
\end{array}
$$
More explicitly
\beq\label{brown1}
{\tau(x)^{Bwn} = \frac{\gamma}{2kT}\left[ (L-\delta)^2 - (L-x)^2  \right] =
\frac{\gamma}{2kT}(x-\delta)(2L-\delta-x)}.
\eeq

\subsection{Model 2: two absorbing boundaries}

\begin{figure}[h!]
\begin{center}
\includegraphics[width=4.6in]{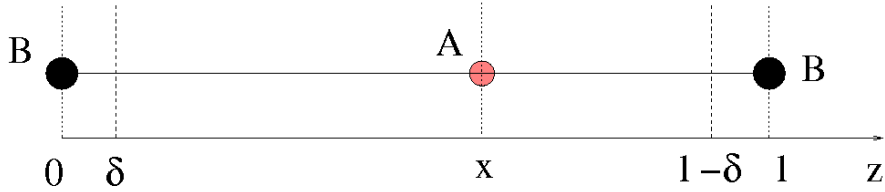}
\end{center}
\caption{\label{initABB}(Color online) A generic initial condition of Model 2 ($t=0$). Two molecules $B$ are fixed
at the boundaries $x=0$ and $x=l$; $x$ and $l-x$ are the initial distances between $A$ and the $B$s and $\delta$ is the distance at which $A$ and $B$ react.}
\end{figure}

Let us now consider the alternative model where two particles $B$ are fixed at positions $z = 0$ and $z = l$,
so that the particle $A$, initially located at $z = x$ (see Figure \ref{initABB}), is absorbed
as soon as it reaches $z = \delta$ or $z = l - \delta$. Such a model is mathematically similar to the previous one.

Simply the deterministic force and the boundary conditions have to be modified. Consequently Eq.(\eqref{eqG}) is to be replaced by

\beq\label{eqG_b}
{ \frac{\partial}{\partial t} G (x,t) = - \frac{1} {\gamma}  \left \{ \left[ F(x) - F(l - x) \right] \frac{\partial}{\partial x}
 G(x,t) - kT \frac{\partial^2}{\partial x^2} G(x,t) \right \}}
\eeq

now $G(x,t)$ is defined as  $ \displaystyle G(x,t) = \int_\delta^{l - \delta}dz\ p (z,t|x,0)$.
The initial condition $p(z,0|x,0) = \delta(x-z)$  gives

\beq
G(x,0) = 1  , \hspace{0.4cm} \text{if} \ \ \delta\le x\le l - \delta \hspace{0.4cm} \text{and} \ \  G(x,0) = 0 ,
\hspace{0.4cm} \text{if not} ,
\eeq
and the absorbing boundary conditions give

\beq\label{absG_b}
G(\delta, t) = 0 \hspace{0.4cm} \text{and} \ \ G(l - \delta, t)  = 0 \hspace{0.4cm}, \forall t>0.
\eeq

The mean first-passage time $\tau(x)$ defined above, then satisfies the ordinary differential equation

$$
- 1 =  \frac{1}{\gamma} \left \{\left[ F(x) - F(l - x) \right]  \frac{d\tau(x)}{dx} - kT \frac{d^2 \tau(x)}{dx^2} \right \},
$$

with boundary conditions

$$
\tau(\delta) = \tau(l - \delta) = 0.
$$

The solution is found to be \cite{Gardiner}

\beq
\begin{array}{r}\nonumber\displaystyle{
\tau(x) = 2 \left\{ \int_\delta^{l-\delta} \,dy\,\frac{1}{\psi(y)} \right\}^{-1} \left\{ \int_\delta^x \,dy\,\frac{1}{\psi(y)}
\int_x^{l-\delta} \,dw\,\frac{1}{\psi(w)} \int_\delta^w \,dz\,\frac{\gamma }{kT} \psi(z) \right.  }
\,\\
\\
\displaystyle{ \left. - \int_x^{l-\delta} \,dy\,\frac{1}{\psi(y)}
\int_\delta^x \,dw\,\frac{1}{\psi(w)} \int_\delta^w \,dz\,\frac{\gamma }{kT} \psi(z) \right\}  }
\end{array}
\eeq

with

\beq
\psi(x)
=  \exp\left\{ \int_\delta^{x} \frac{F(s) - F(l-s)}{kT} \ ds\right\}=
\exp\left\{ - \frac{U(x) - U(l-x) - U(\delta) +U(l-\delta)}{kT} \right\}
\eeq

\medskip

since $F(x) = -{\partial U(x)}/{\partial x}$. After simplification, one has

\beq \label{Tx_b}
\begin{array}{l}
\displaystyle{ \tau(x) = \frac{\gamma}{kT} \left\{ \int_\delta^{l-\delta} \ dy \ \phi(y) \right\}^{-1}
\int_\delta^x  dy  \int_x^{l-\delta}  dw  \int_y^w  dz  \frac{\phi(y)  \phi(x)}{\phi(z)} }
\end{array}
\eeq
where
\[
\phi(s) = \exp\left\{\,- \frac{U(s) -U(l-s)}{kT} \right\}\ .
\]

Similarly to the Model 1, the expression for the mean reaction time with Brownian motion only ($U=0$),
is particularly simple

\beq\label{brown2}
\tau(x)^{Bwn} = \frac{\gamma}{2kT}\,(x-\delta)(l-\delta-x) \ .
\eeq

To summarize, in this Section we have obtained the general form of the mean first-passage time $\tau(x)$, that is the average time needed by molecule $A$ to reach the molecule $B$ (or one molecule $B$ in the case of Model 2), as a function of the initial intermolecular  distance $x$ and temperature $T$, for both Model 1 [Eq.(\ref{Tx})] and Model 2 [Eq.(\ref{Tx_b})], respectively.
The same function is given for randomly driven encounters between the reaction partners in Eq.(\ref{brown1}) for Model 1, and in Eq.(\ref{brown2}) for Model 2.

\section{Quantitative theoretical predictions}\label{applications}

In order to answer the question of whether it would be feasible to {\it experimentally detect} the possible existence of a deterministic attractive force through which the cognate partners of biochemical reactions interact at long distance, we first have to delimit the physical context, choose the domain of physical parameters and provide the analytic form of the two-body interaction potential. In what follows, a long range resonant potential potential $U(x)=-{\cal C}/x^3$ is considered. As discussed in the Introduction, this kind of interaction can have sizeable effects at long distances at
variance with London - Van der Waals $1/x^6$ interactions (see also the Appendix).
Following Fr\"ohlich \cite{frohlich4}, a lower bound for the coefficient of this
potential is given by ${\cal C} \simeq \hbar e^2 (Z_A Z_B)^{1/2}/2 M\omega_0 \varepsilon^\prime(\omega_0)$, where
$Z_A$ and $Z_B$ denote the number of charges of averaged mass $M$ and charge $e$ contributing to the
dipole moment of the molecules $A$ and $B$ respectively, whereas $\omega_0$ stands for their
oscillation frequency; $\varepsilon^\prime(\omega_0)$ is the real part of the dielectric
constant of the interposed medium. In particular, it is interesting to remark that in the expected range of oscillation frequencies for the setup of collective dipole oscillations in macromolecules (that Fr\"ohlich
estimated to be $\omega_0 \simeq 10^{11} - 10^{12} \mathrm{Hz}$) the value of $\varepsilon^\prime(\omega_0)$ drops down to a few units \cite{ellison} thus allowing a much smaller screening of the interactions with respect to the static case. In this context, we use $Z_A = Z_B = 1000$ (see Ref. \cite{pokornybook}) and
the proton mass for $M$. A convenient unit system remains to be chosen. For the numerical tabulations of $\tau (x)$, instead of c.g.s. units we use $\mu$m, kDa, and $\mu$s, with the following definitions $1$ $\mu$m =$10^{-4}$cm, $1$ kDa = $10^{-21}$g and $1$ $\mu$s = $10^{-6}$s.

In this system of units
we evaluate the lower bound of $\mathcal{C}$ which is found to be $\mathcal{C}\simeq 10^{-30} \mathrm{erg.cm^3} = 0.1 \ \mathrm{kDa.\mu m^5.\mu s^{-2}}$.
Henceforth, we shall consider ${\cal C}$ varying from $0.1$ to $10\ \mathrm{kDa.\mu m^5.\mu s^{-2}}$.
These values are given with a degree of arbitrariness
that can be reduced by considering that ${\cal C} = 10$ corresponds to the physical situation where $U(x)\simeq kT$ at
$x\sim 0.1 \mathrm{\mu{m}}$.  Hence the choice ${\cal C}\in [0.1, 10]$
is a very cautious estimate with respect to those existing in the literature about a possibly larger range of action
(it has been surmised by Fr\"ohlich and others that $U(x)$ might become comparable with $kT$ at $x\sim 1 \mu{\rm m}$ or
more \cite{pokornybook, rowlands,tuszynski}). Among other constants appearing in equations \eqref{Tx} and \eqref{Tx_b},
the friction coefficient $\gamma$ of the molecule $A$ has been estimated according to Stokes' law
 $\gamma = 6 \pi \eta(T) \mathcal{R}$, where $\eta(T)$
corresponds to the viscosity of water at temperature $T$ and ${\cal R}$ stands for
the hydrodynamic radius of the molecule. The value of $\mathcal{R}$ has been set equal to
$ 5\cdot 10^{-3} \mathrm{\mu{m}}$
which is the typical diameter of a biomolecule with a mass in the interval $50 - 100 \mathrm{kDa}$
(proteins and DNA fragments); the same value has been fixed for the reaction radius $\delta$ introduced
in both models : ${\cal R} = \delta = 5\cdot 10^{-3} \mathrm{\mu{m}}$.

All the computations of $\tau(x)$
have been performed by means of MATLAB programs. Also, as MATLAB does not allow to perform direct integrations
over non-rectangular domains, integrals with variable limits in equations \eqref{Tx} and \eqref{Tx_b} have
been first ``vectorized'' \cite{shure} for each $x$ to calculate $\tau(x)$ with a recursive adaptive Simpson quadrature
(MATLAB \texttt{quadl} function). Further checks on the reliability of the method have been done through  direct numerical integration of the Langevin equation \eqref{lang_adiab} by means of a standard Euler-Heun algorithm and by averaging over $10^4$ different
realizations of the random walk. Minor precision problems especially when $x \sim \delta$ have been thus detected and corrected in what follows.

\subsection{Model 1}

\begin{figure}
\begin{center}
\includegraphics[width=4.6in]{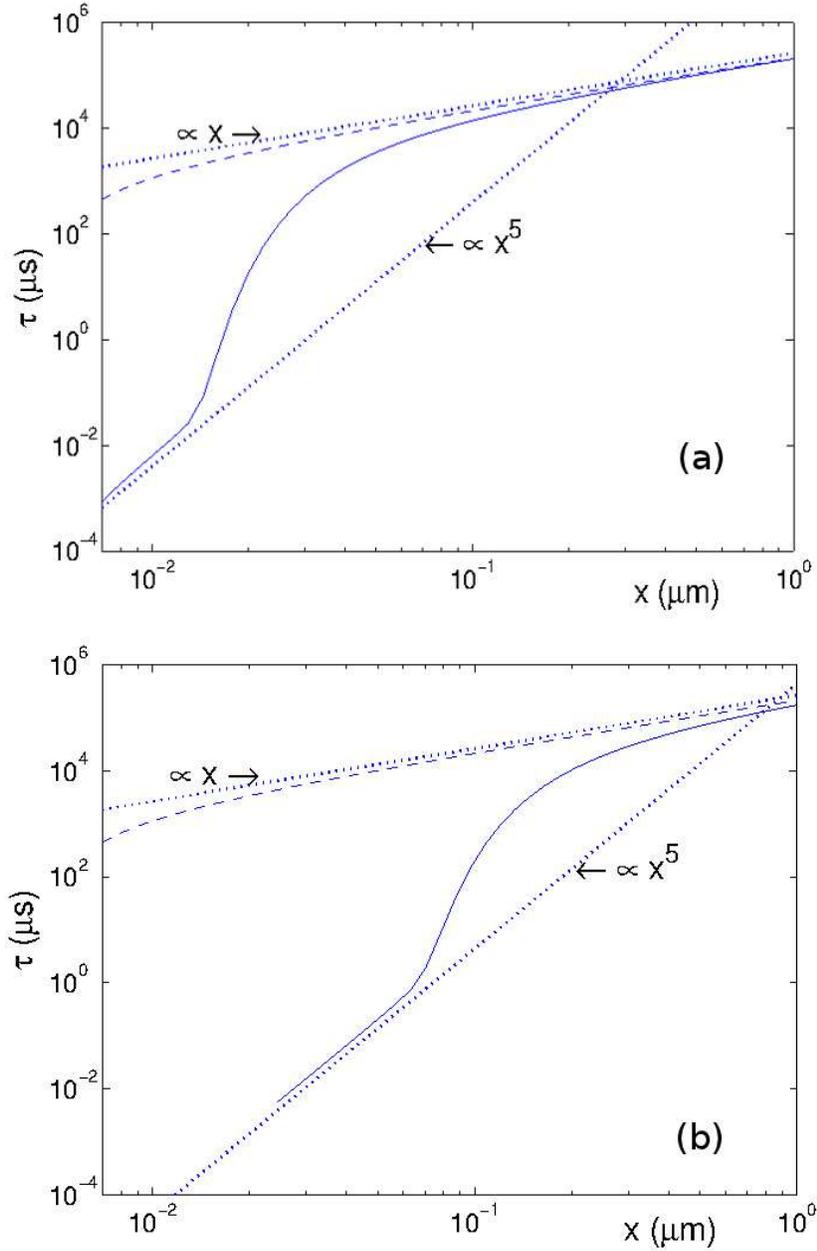}
\end{center}
\caption{(Color online) Model 1: mean encounter time $\tau$ between two molecules $A$ and $B$, initially placed at a distance $x$ one from the other. Dotted lines are asymptotic behaviors. Dashed line refers to purely random encounters.
Solid line refers to the combined effect of a random force plus a deterministic one derived from the potential $U(x)=-{\cal C}/x^3$. Figure (a) refers to ${\cal C}=0.1$. Figure (b) refers to ${\cal C}=10.$}
\label{FigC01A}
\end{figure}

We have computed $\tau (x)$ and $\tau (x)^{Bwn}$ by means of Eqs.(\ref{Tx}) and (\ref{brown1}) respectively,
where we have set $R=\delta=5 \cdot 10^{-3}  \mathrm{\mu m}$ and $U(x) = - \mathcal{C}/x^3$, as detailed
above. The position $L$ of the reflecting barrier characteristic of Model 1 has been fixed so as $x\ll L$ for all $x$. In particular, $L=10 \mathrm{\mu{m}}$ and a maximal value for $x$ equal to $1 \mathrm{\mu m}$ have
been used. Figure \ref{FigC01A} displays the numerically found shapes of both functions $\tau (x)$ and $\tau (x)^{Bwn}$ computed
at $T= 300 \mathrm{K}$ and for different values of the attractive potential coefficient $\mathcal{C}$. A first check on the reliability of the plotted results is done by observing that $\tau(x)< \tau(x)^{Bwn}$ for all the $x$ values while both curves merge at large $x$ values, as expected when the resonant attraction is
wiped out by thermal noise. In particular, as we always considered $x\ll L$, the asymptotic behavior of $\tau(x)^{Bwn}$ is then proportional to $x$ as required by  Eq.(\ref{brown1}). On the contrary, at smaller $x$,
$\tau(x)^{Bwn}$  bends downwards to slightly smaller values with respect to the extrapolated linear
dependence (this happens when $x$ is no longer much larger than $\delta$). At variance, the pattern of $\tau(x)$ has two asymptotic limiting behaviors: at large $x$ values it joins the Brownian curve $\tau(x)^{Bwn}$, and at small $x$ values it is $\tau(x)\sim x^5$ (a power-law characteristic of the $1/x^3$ form of the potential); the latter might be anticipated on the basis of simple dimensional arguments since the l.h.s. of Eq.\eqref{lang_adiab} has the
dimensions $[dz/dt] = l t^{-1}$ while the r.h.s. leads to $[{\cal C}/x^4] =[{\cal C}] l^{-4}$ in a purely
deterministic regime. By combining the two, for
a generic time scale $[\tau]=t$ associated with a displacement length
$x$ we get $\tau \sim [x]^5$ \cite{nota2}. The two limiting behaviors are bridged by a steep transition pattern which moves rightward or leftward according to the value of $\mathcal{C}$, as shown on Figure \ref{FigC01A};
the stronger the potential the larger the $x$-values at which
$\tau (x)$ displays the knee joining the $x^5$ functional dependence. The transition pattern of $\tau(x)$ is steep since the reflecting barrier is located far from the only one molecule $B$.

\begin{figure}
\begin{center}
\includegraphics[width=4.6in]{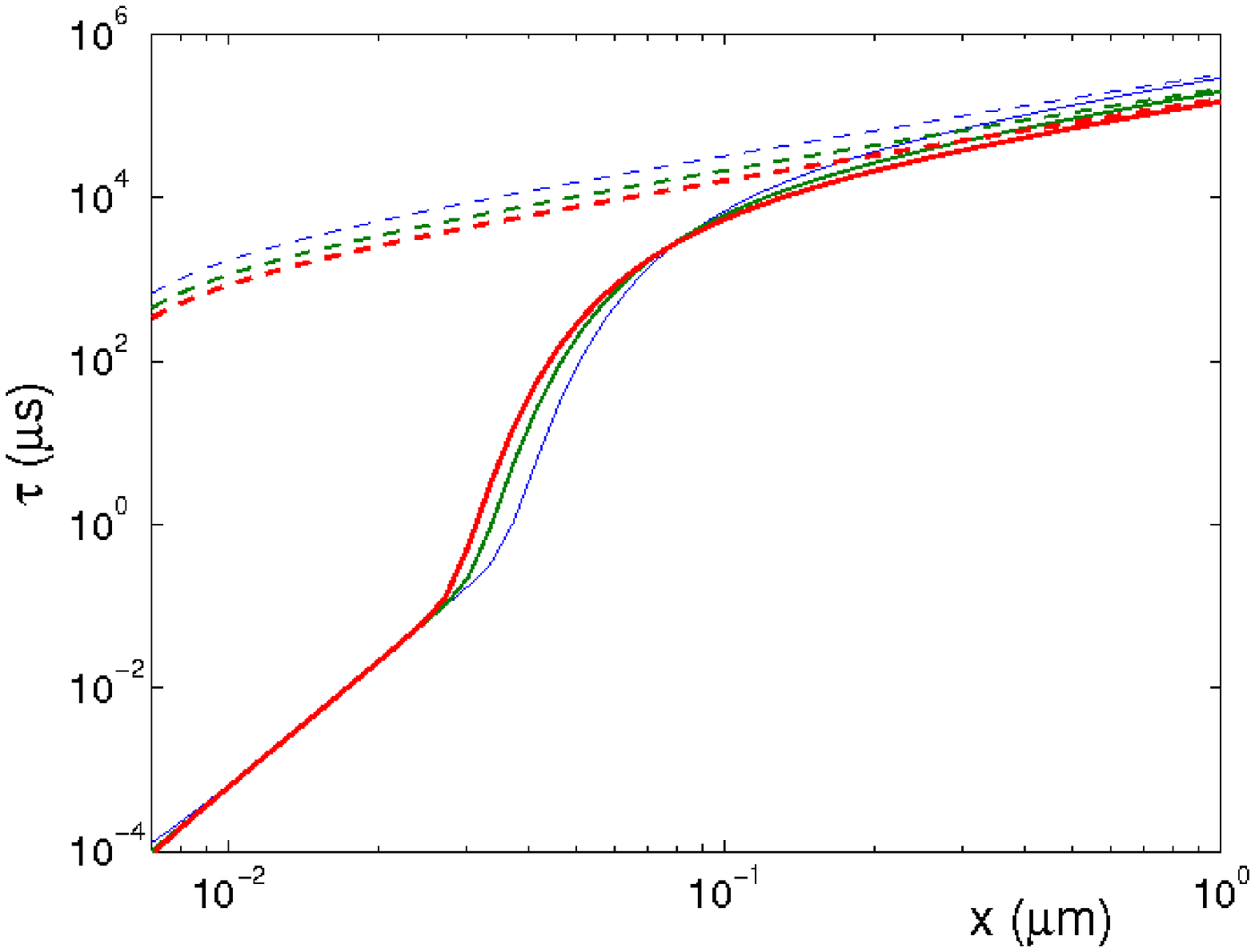}
 \end{center}
\caption{(Color) Model 1: temperature dependence of $\tau (x)$ for ${\cal C}= 1.0$. Red solid and dashed curves refer to $T=200\text{K}$, green dashed and solid curves refer to $T=300\text{K}$, blue dashed and solid curves refer to $T=400\text{K}$.}
\label{FigTempA}
\end{figure}

In any case, $\tau (x)^{Bwn}$ is found to exceed $\tau (x)$ by a factor of, say, $10$ at definitely larger
$x$-values and, what is more relevant, at longer values of first encounter time : with ${\cal C}=0.1$,
such a difference occurs at $x\simeq 300 \mathrm{\AA}$ where $\tau^{Bwn}\simeq 6 \mathrm{m s}$ and $\tau\simeq 600 \mu s$, while with
${\cal C}=1.0$, it occurs at $x\simeq 640 \mathrm{\AA}$ where $\tau^{Bwn}\simeq 10 \mathrm{m s}$ and $\tau\simeq 1 \mathrm{m s}$, and
with ${\cal C}=10$ at $x\simeq 1400 \mathrm{\AA}$ where $\tau^{Bwn}\simeq 20 \mathrm{m s}$ and $\tau\simeq 2 \mathrm{m s}$. As we
shall see in the next
Section, should we interpret $x$ as the average distance between any two reacting molecules in three dimensions, this range of
$x$-values (between a few hundreds Angstroms and $1 \mu{\rm m}$) is easily attained by varying the concentrations of the
reactants between a few micro-Moles down to one nano-Mole. Notably, the encounter times belong to an interval of values easily accessible by means of optical detection methods.

A priori, further qualitative indications on the possible presence of attractive deterministic forces between
cognate partners could be observed by modifying the temperature of the system. As shown in Figure \ref{FigTempA}, $\tau (x)$ and $\tau(x)^{Bwn}$
plotted  for three different values of $T$ confirm that the $x^5$ functional
dependence is purely deterministic as $T$ has no influence within this domain.
On the contrary, $\tau(x)^{Bwn}$ displays the same dependence on $T$ for all values of $x$ . Surprisingly, the steep transition pattern of $\tau(x)$ at
intermediate values of $x$ is characterized by a temperature dependence which is inverted compared to the Brownian case : in presence of an intermolecular potential, the higher the temperature the larger the first-passage time of $A$ at $x=\delta$. Finally note that the temperature range considered in Figure \ref{FigTempA} is a broad one ($T = 200, \ 300,$ and $ 400  \mathrm{K}$).
Nevertheless,
as the physiological temperature range corresponds only to a few percent around $300 \mathrm{K}$, it is likely that
variations of the first passage time at different temperatures are too weak to be experimentally detectable within such an interval. In particular, computations performed for temperature
differences of $10 \mathrm{K}$ (typically $290, 300, 310 \mathrm{K})$ with the Model 1 show variations
of $\tau$ less than five percent of its value in the Brownian case as well as in the case of Brownian plus deterministic force.

\subsection{Model 2}

\begin{figure}[h!]
\begin{center}
\includegraphics[width=4.6in]{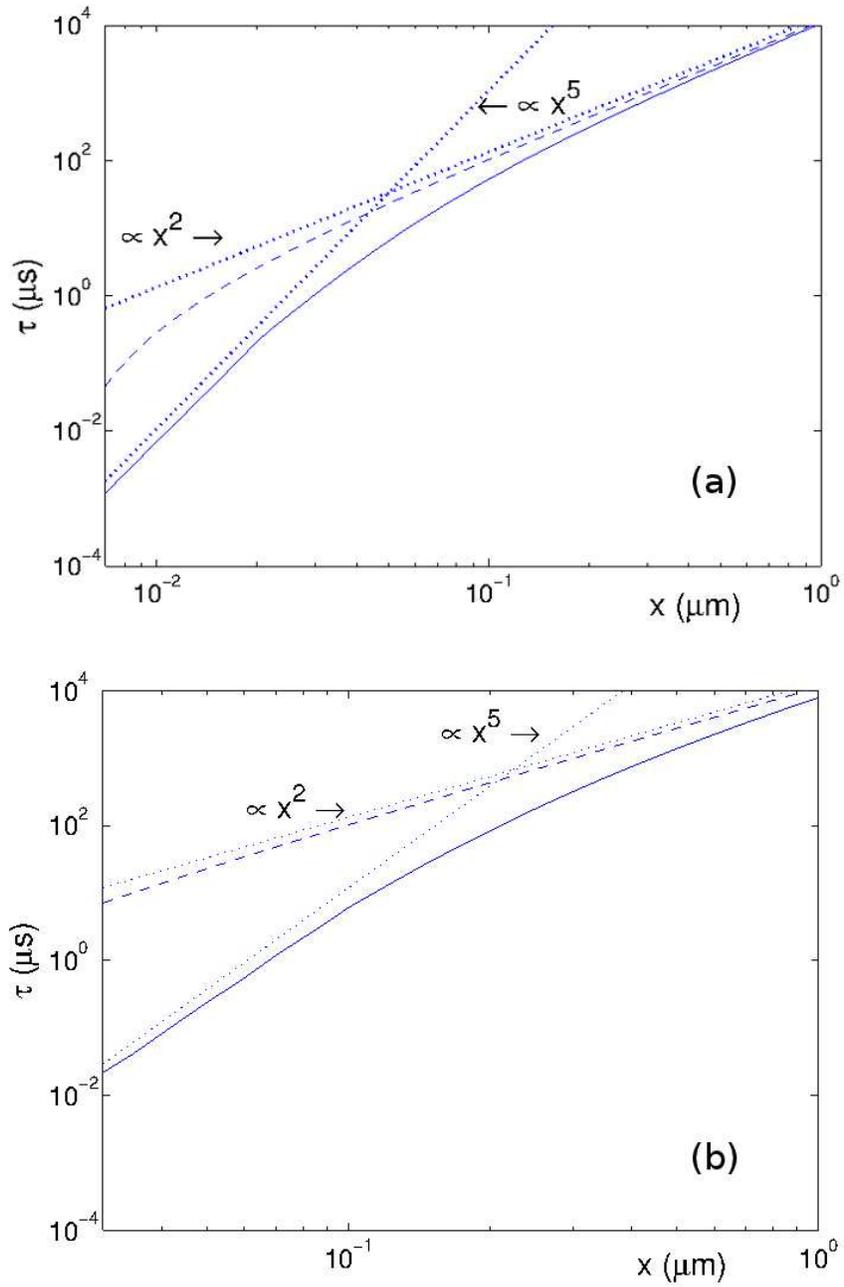}
\end{center}
\caption{(Color online) Model 2: mean encounter time $\tau$ between two molecules $A$ and $B$, initially placed at a distance $x$ from each other.
Dotted lines are asymptotic behaviors. Dashed line refers to purely random encounters.
Solid line refers to the combined effect of a random force plus a deterministic one derived from the potential $U(x)=-{\cal C}/x^3$. Figure (a) refers to ${\cal C}=0.1$. Figure (b) refers to ${\cal C}=10.$.}
 \label{FigC01B}
\end{figure}

We have also plotted $\tau (x)$ and $\tau (x)^{Bwn}$ computed according to Eqs.(\ref{Tx_b}) and (\ref{brown2}) respectively,
where $R=\delta=5 \cdot 10^{-3}  \mathrm{\mu m}$, $l = 2 x$, and $U(x) = - \mathcal{C}/x^3$ as in the case of Model 1.
Figure \ref{FigC01B} displays the numerically found shapes of both functions $\tau (x)$ and $\tau (x)^{Bwn}$ computed again
at $T= 300 \mathrm{K}$ and for different values of the attractive potential coefficient $\mathcal{C}$.

\begin{figure}[h!]
\begin{center}
\includegraphics[width=4.6in]{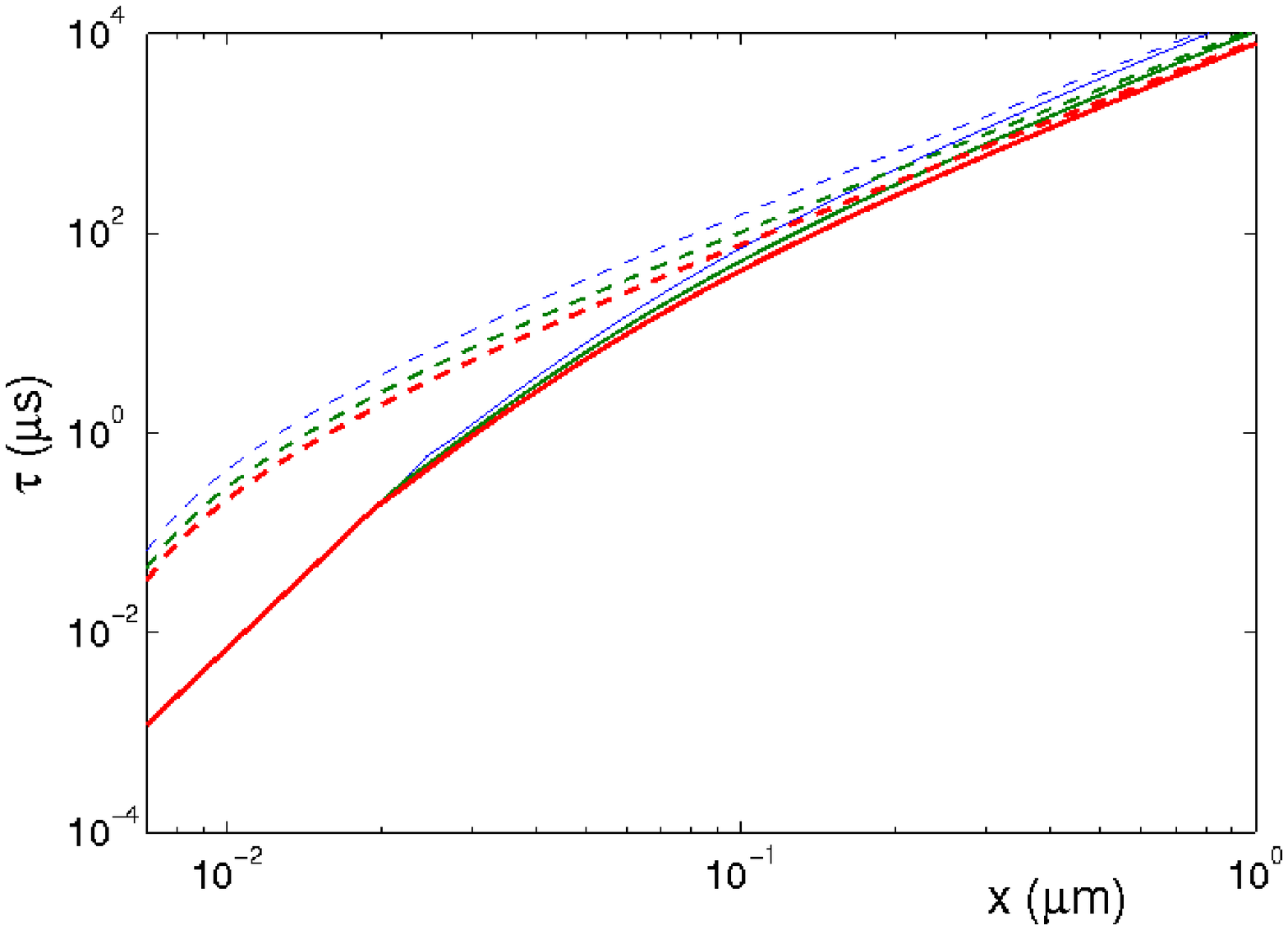}
 \end{center}
\caption{(Color) Model 2: temperature dependence of $\tau (x)$ for ${\cal C}= 0.1$. Red solid and dashed curves refer to $T=200\text{K}$, green dashed and solid curves refer to $T=300\text{K}$, blue dashed and solid curves refer to $T=400\text{K}$.}
\label{FigTempB}
\end{figure}

At first check, the characteristics of $\tau(x)$ prevailing for Model 1  (i.e., the $x^5$
functional dependence at small $x$, and the tendency of $\tau(x)$ to join $\tau(x)^{Bwn}$ at large $x$) also apply here.
Nevertheless, the asymptotic $x$-dependence of $\tau(x)^{Bwn}$ is now proportional to $x^2$ as obtained from
equation Eq.(\ref{brown2}) by replacing $l$ by $2 x$ when $x\gg \delta$. On the other hand, the absence of a reflecting
barrier makes the steep transition feature for $\tau(x)$ disappear and be replaced by a mild crossover at
intermediate values of $x$.
This steeper pattern for the first passage time as a function of $x$ - in the case of purely Brownian diffusion - entails a
less pronounced separation between $\tau(x)$ and $\tau (x)^{Bwn}$. In fact, a separation between these two curves by a
factor of 10, (to make the same kind of comparison that we did for Model 1) occurs for ${\cal C}=0.1$ at $x\simeq 200 \mathrm{\AA}$
for which $\tau^{Bwn}\simeq 1.1 \mathrm{\mu s}$ and $\tau\simeq 0.1 \mathrm{\mu s}$; while for ${\cal C}=1.0$ this
happens at $x\simeq 600 \mathrm{\AA}$
for which $\tau^{Bwn}\simeq 30 \mathrm{\mu s}$ and $\tau\simeq 3 \mathrm{\mu s}$; and, finally, for ${\cal C}=10$ at
$x\simeq 1170 \mathrm{\AA}$
for which $\tau^{Bwn}\simeq 120 \mathrm{\mu s}$ and $\tau\simeq 12 \mathrm{\mu s}$. Therefore, we can see that Model 2 is more
constraining than Model 1, in the sense that, at equal values of $\cal C$ (that is, at equal strength of the
long range interaction) smaller intermolecular distances and a much faster tracking of the dynamics of the
reactants are needed to discriminate with the same degree of confidence (arbitrarily set as
a factor of 10) between random and non random encounters of the reactant molecules.

The temperature dependence of both $\tau (x)^{Bwn}$ and $\tau (x)$ is reported in Figure \ref{FigTempB}.
The main features of $\tau(x)$ in Model 1 are likewise present in Model 2. In addition, we can see
that the inversion in the temperature dependence with respect to the Brownian case (which was characteristic of the steep transition pattern of Model 1) is no longer there. Thus, the peculiar temperature dependence of this steep transition pattern could be mostly attributed to the presence of the reflecting barrier
characteristic of Model 1. Likewise in Model 1, we used again $T= 200,\ 300$ and $400 \mathrm{K}$, even though computations carried out at physiological temperature again yield too weak variations of $\tau(x)$ to be experimentally detectable.

\section{Discussion and concluding remarks}
\label{conclusions}
The numerical results reported in the preceding Section are in favor of a  positive answer to the main
question addressed by the present work.
In fact, the numerical study of Models 1 and 2  revealed qualitative differences in the
mean first passage time $\tau$ between the case of a pure Brownian diffusion of the molecule $A$ (see Section \ref{reactions}) and the case
in which an attractive (resonant) potential $U$ is added to a random force. In particular, in the latter case, the functional dependence
of $\tau$ on the initial distance $x$ between the molecule $A$ and its target (molecule(s) $B$) demonstrates
the existence of different patterns in the two models depending on the range of the $x$ values considered :

\begin{itemize}
 \item a deterministic pattern at small $x$ values (small initial separations), characterized
by a power law representative of the potential under consideration ($x^5$ for the resonant potential used, $x^{p+2}$ for a general
potential of the form $U(x) \propto {x^{-p}}$ );
\item a Brownian pattern at large initial separations, proportional to $x$ or to $x^2$ depending upon the symmetry of
the system ($x$ for an asymmetric situation as described by Model 1, and $x^2$ for a symmetric one as described by Model 2);
\item a steep transition pattern joining the two asymptotic ones in the case of Model 1, and a smooth crossover joining the two asymptotic ones in the case of Model 2.
\end{itemize}

Although complementary computations revealed some interesting features in the temperature dependence of
$\tau$, the corresponding degree of variation in a laboratory accessible interval of temperature is too weak to be experimentally detectable.

In any case, it is obvious that $x$ must constitute an experimentally accessible control parameter so that the results mentioned above may be used to predict the possible role of long-range intermolecular
forces in biological processes. Notably, such an approach is not so usual. Indeed, most of the attempts made hitherto in this direction have resulted in experimental measurements of association constants $k_a$ (characteristic of a reaction medium),
which are predictable from the Smoluchowski theory also when intermolecular forces are considered \cite{debye,noyes}.
The focus of Smoluchovski theory is on the association constant $k_a$ which represents the probability for two molecules to react per time unit, irrespective of their position.
In the case of Brownian encounters, this is given by \cite{hippel2} $\ k_a^B = 4 \pi R D \equiv 4 \pi\delta k T /\gamma$ where $R$ is the reaction radius that can be approximated to $\delta$ from the current study, and $D$ is the sum of the diffusion coefficients of the two cognate partners. In the presence of some interaction potential $U$, one has
$k_a^*=4 \pi R^* D$ where $R$ has been replaced by \cite{hippel2,debye,noyes} $\ \textstyle{R^*= R \left(\int \limits_R^\infty r^{-2} e^{-U(r)/k T} dr\right)^{-1}}$.
Now, if $k_a$ is experimentally measured for some reaction and it turns out that $k_a^B <k_a$, then this would indicate that some deterministic force is
in action but one can hardly find out the law of the interaction potential because after integration over $r$ there is no one-to-one
correspondence between $R^*$ (thus $k_a^*$) and the functional form of $U(r)$ \cite{tuszynski}. On the other hand, in measuring  $k_a< k_a^B$, we cannot be sure that the reaction is simply diffusion-driven because, in this case, chemical times could be long
enough to make $k_a$ smaller than the corresponding Brownian value. The advantage of our dynamical approach is that our
models still apply by choosing $\delta$ as the distance at which $A$ and $B$ get in contact without reacting, and with the experimental technique discussed below (FCCS) we can make a distinction between the association time and the chemical times.

In the present situation $x$ values might to some extent be considered in three dimension as the {\it average distance} between two molecular partners $A$ and $B$, while this quantity can be easily controlled in laboratory experiments by varying the concentrations of the reactants. Given the concentrations ${ C}_A=N_A/V_A$ and
${ C}_B=N_B/V_B$ (with $V_{A,B}$ = the initial volumes and $N_{A,B}$ = the number of molecules of the two
species respectively; remark that these numbers are controlled through the molarity, i.e. a definite fraction of the Avogadro number), we get the estimate $x={ C}_{av}^{-1/3}$ for the average intermolecular distance
from the average concentration ${ C}_{av}=(N_A + N_B)/V$, where the reaction volume $V=V_A+V_B$.  In practice, as an
example, with ${ C}_{av}=1 \mathrm{nM}$ we have $x\simeq 1 \mathrm{\mu m}$ as the average distance between any two molecules, while with ${ C}_{av}= 1 \mathrm{\mu M}$ we have $x\simeq 1000 \mathrm{\AA}$. By working at equimolarity, that is ${ C}_A={ C}_B$, then ${ C}_{av}^{-1/3}$
is a good estimate of the average distance between one $A$ and one $B$ molecule. Working with nano-Moles of DNA and
proteins (enzymes, transcription factors) is quite
standard in molecular biology experiments.  With such concentrations of reactants, both models (1 and 2)
predict that the first passage time -- that can be interpreted as the average encounter time
between one $A$ and one $B$ molecule initially located at intermolecular distances of a few thousands of
Angstroms -- varies in the interval between a few tens of microseconds to
about one millisecond in the presence of an attractive deterministic force that would sum up to the random force. On the contrary, in the very same conditions, random only driven encounters would exceed the above mentioned encounter times by one or two orders of magnitude.
Again, the distance at which sizeable differences could be observed may vary significantly depending on the actual value of the resonant potential parameter ${\cal C}$.
On the other hand, estimates in literature \cite{rowlands,tuszynski,pokornybook} suggest that these long-range resonance
interactions could be effective up to distances in the order of $1 \mu{\rm m}$ (the action range is  estimated
by computing the distance at which the resonance interaction energy equals the level of thermal noise $kT$).
In this respect, in the preceding Section we have limited ourselves to cautious
estimates for the parameter ${\cal C}$, focusing on conservative assumptions for average encounter time
varying in the interval $10^{-5} - 10^{-3}$ seconds which can be readily detected with the aid of Fluorescence
Cross-Correlation Spectroscopy (FCCS technology).
This is a powerful technique which is being increasingly applied to the study of diffusion and chemical reaction rates  in complex biological systems using fluorescently labeled macromolecules \cite{maiti,hom,schwille}.
FCCS measures the spontaneous fluctuations
of fluorescences $\delta F_1(t)$ and $\delta F_2(t)$  that arise from the diffusion of fluorescently labeled molecules
of type $1$ and $2$, respectively - illuminated by two laser light beams of different colors - into or out of an open sampling volume. Even though the size of the detection
volume is diffraction limited, the autocorrelation functions of $\delta F_{1,2}(t)$ and the cross-correlation function $\langle\delta F_1(t)\delta F_2(t)\rangle$ can be altered by processes
occurring on smaller spatial scales. These correlation functions provide information on diffusion
properties of fluorescent molecules.

Of course, we are well aware of the fact that the models studied here are simplified descriptions of the reality. Indeed, protein-protein and protein-nucleic acid interactions in vivo generally take place within complex structural scaffolds such as the membrane cytoskeleton or the chromatin envelope, which are themselves the subject of highly dynamical regulations (e.g., Refs. \cite{ww,xx}); and may also possibly interfere with the spatiotemporal control of the given reactions (e.g., Refs. \cite{yy,zz}). Should resonant electrodynamic interactions be involved within such an intricate context, it seems illusory at this stage to assess realistic ${\cal C}$ values simply based on the proposed experiments. In fact, regarding protein-DNA in vivo (physiological) interactions for instance, it may well be that the putative values fluctuate depending on a host of variables, possibly including - in a non-mutually exclusive way, charges on proteins and DNA, the effect of surrounding electrolytes, the nucleic-/amino-acid compositions, the length of accessible DNA, etc.
However, we stress here that our initial goal, as described in this article, is to merely probe whether or not biological partners can take advantage, besides thermic diffusion, also of long-distance (0.1-1$\mu$m) forces of electrodynamic origin to eventually interact. If established, this novel concept would then in turn open new avenues of research to investigate long-standing biological issues, e.g., on the precise definition of which variables exactly pertain on protein-DNA interactions, and how a diffusing protein particle may actually recognize the particular cognate DNA site among many other locations also available.
Since we do not expect dramatic qualitative changes out of the numerical simulations of Eqs.(\ref{mol_dyn-full}) in
three dimensions \cite{seshadri}, an experimental setup providing a practical realisation of what has been investigated in
the present work could be devised by resorting - as experimental probes - to three broad classes of interactions: protein-DNA, protein-RNA, and protein-protein (ligand-receptor). As DNA and RNA molecules have not a preassigned length, it is implicitly understood that only short fragments are to be considered (some tens or a few hundreds of base pairs, that is, oligonucleotides or plasmides respectively). The proteins interacting with DNA or RNA can be processing enzymes (helicases, polymerases, recombinases) or transcription factors normally bound at promoters, enhancers, insulators, or silencers. Thus, for example, one could choose two molecular species consisting, respectively, of a short double stranded DNA molecule (for example a synthetic oligonucteotide of $\sim 100$ base pairs or even less) and a protein with a
site specific affinity for the chosen DNA molecule (i.e., a transcription factor).
By labeling the DNA molecules and proteins with standard fluorophores their dynamical behavior can be followed by means of FCCS microscopy at different concentrations ${ C}={ C}_A={ C}_B$
of the reactants to get a characteristic time scale as a function of $x={ C}^{-1/3}$. In this way such an
experimental
set up should provide - after data fitting -  an estimate of the constant ${\cal C}$ for the resonant
potential considered above. Thus, ${\cal C}=0$ would mean that the reactants meet only under the action of Brownian diffusion,
whereas ${\cal C}\neq 0$ would prove the existence at the same time of the long-range interactions evoked
throughout this paper and give quantitative information about them.

\acknowledgments

We warmly thank V. Calandrini, R. Lima, S. Jaeger and D. Marguet for many fruitful discussions.
Work of the M.P. group has been supported by a BQR grant of the former University of Aix-Marseille II and
by a PEPS grant of the CNRS.
Work in the P.F. laboratory is supported by institutional grants from INSERM and CNRS, and by specific grants from the "Fondation Princesse Grace de Monaco", the "Fondation de France", the "Association pour la Recherche sur le Cancer" (ARC), the "Fondation pour la Recherche Médicale" (FRM), the "Agence Nationale de la Recherche" (ANR), the "Institut National du Cancer" (INCa), and the Commission of the European Communities.

\section{appendix}
For the sake of clarity and to help the reader to get a hold of the physical origin of the $U(R)\propto - 1/R^3$ potential
referred to throughout the present work, this Appendix provides some theoretical elements about the interaction of oscillating electric dipoles.

To begin with, let us recall some basic fact on this subject. Two atoms (or two small molecules) $A$ and $B$ in their ground states with no net charge excess and vanishing average dipole moment (i.e. both are nonpolar) interact through the London - Van der Waals dispersive force. The origin of this interaction is as follows.
Though the expectation values of the dipole operators are zero for nonpolar atoms, quantum fluctuations are responsible
for their instantaneous non vanishing dipolar moments. This entails a non zero dispersion of dipole moment operator. The energy of the two isolated atoms is corrected at first order by the dipole-dipole interaction energy which is proportional to the average dipole moments, thus it vanishes when both atoms are in their ground states. Instead, the second order perturbative correction, due to the coupling between instantaneous dipole fluctuations, is found to be proportional to $1/R^6$. (In a QED framework the London - Van der Waals interaction stems from the exchange of virtual photons between the atoms). This is a short range potential, so called because the
exponent of the power law of $R$ is strictly larger than $3$, the dimension of physical space. London - Van der Waals interactions are of generically weak intensity, whereas they likely become of prime importance in a biological context when acting at short distances (below the Debye length) together with additional interactions of chemical type \cite{israel}.

Remarkably, the first order perturbative correction may be non-vanishing under a degeneracy condition. Indeed, if one or both atoms are in an excited state, provided that the condition for exchange symmetry is fulfilled, that is, they have
common eigen-energies in their spectra, it can be shown \cite{margenau} that the interaction energy is now proportional to $1/R^3$, a long range potential.

Interactions of similar kinds to those just mentioned between two atoms (or small molecules) could exist between macromolecules with an oscillating electric dipole moment. In this case, the oscillating dipole moment would not be due to
the electron motions but, rather, to conformational vibrations. As already mentioned in the Introduction, this interaction
between the oscillating electric dipole moments of reacting macromolecules could play a relevant role in living matter.
In fact, as already quickly recalled in the Introduction, the high static dielectric constant of water together with the considerable amount of ions present in living cells tend to screen any electrostatic interaction beyond a distance of a few
Angstroms. However, this electrostatic opaqueness does not hold for an oscillating field: the higher the frequency of an
oscillating field the more transparent an aqueous salted medium. In fact, the value of the dielectric constant of water at room temperature is a decreasing function of the frequency \cite{ellison} and, for example, already at 1THz $\varepsilon(\omega =10^{12})\simeq 4$; likewise, the imaginary (dissipative) part of the dielectric constant (which is proportional to the conductivity of the medium due to the presence of free ions) is inversely proportional to the frequency of the oscillating electric field (according to the Drude equation \cite{jackson}), so that at suitably high frequency can be negligible.

Let us a now study the basic mechanism of interaction between two oscillating electric dipoles before discussing its application to biomolecules \cite{pretopettini}.
As we show below, these oscillating dipoles can activate long-range forces that will be shown to be frequency selective. We consider a one dimensional simplified model in which the dipoles oscillate at frequencies $\omega_A$ and $\omega_B$ respectively.
Then a computation of the interaction energy between $A$ and $B$ can be given which, despite the simplified treatment, allows to grasp some basic physical facts.

Let $\mu_A$ and $\mu_B$ be the masses of the two oscillators, let their dipole moments be parallel and given by
$q Z_A r_A$ and $q Z_B r_B$, and assume that their mutual separation $R$ is such that $R \gg r_A, r_B$, then we can write
the interaction Hamiltonian as
\begin{equation}
H = \frac{p_A^{\ 2}}{2 \mu_A}  + \frac{p_B^{\ 2}}{2 \mu_B}  + \frac{1}{2}\mu_A\omega_A^{2} {r}_A^{\ 2}  +
\frac{1}{2}\mu_B\omega_B^{2} {r}_B^{\ 2}+  \frac{\zeta q^2 Z_A Z_B }{4\pi \varepsilon_0 R^3}
{r}_A \ {r}_B,
\end{equation}
where $Z_i$, $i=A,B$, stands for an effective number of charges which account for the average value of the dipole moment
of the oscillator $i$; $\zeta$ is a geometrical factor depending on the orientation of the dipoles with respect to the
line joining them (on which the distance $R$ is measured).
Then, introducing a mean mass $M$ defined so that $ \mu_A = M Z_A $ and $\mu_B = M Z_B$, the Hamiltonian becomes
\begin{equation}
H = \frac{1}{2M} \left( p_A^{\ 2}  + p_B^{\ 2} \right) + \frac{1}{2}M\omega_A^{2} {r}_A^{\ 2}  +
\frac{1}{2}M\omega_B^{2} {r}_B^{\ 2} +  \frac{\beta }{R^3}  {r}_A \ {r}_B,
\end{equation}
where the transformations $(Z_i)^{1/2} r_i   \rightarrow r_i$ and $p_i / (Z_i)^{1/2} \rightarrow p_i$, $i=A,B$, have been introduced
(the variables $r_i$ and $p_i$ are still canonically conjugated) and we put $\beta = \zeta q^2 (Z_A Z_B)^{1/2}/4\pi \varepsilon_0$, where $\varepsilon_0$ is the dielectric constant of vacuum - in the absence of a material medium between
the oscillators - to be replaced by  $\varepsilon(\omega)$ when a medium is present.
In matrix form this also reads
\begin{equation}
H = \frac{1}{2M} \left( p_A^{\ 2}  + p_B^{\ 2} \right) + \frac{1}{2}M
  \left ( {r}_A  \ \ {r}_B \right)
\underbrace{\left( \begin{array}{cc}
\omega_A^{\ 2} & \beta/M R^3   \\
& \\
\beta/M R^3 & \omega_B^{\ 2}  \end{array} \right)}_{C}  \left( \begin{array}{c}
 {r}_A  \\
 \\
{r}_B  \end{array} \right).
\end{equation}
Matrix $C$ is real and symmetric, thus diagonalizable by means of an orthogonal transformation. Let $\omega_+^2$ and $\omega_-^2$ the eigenvalues of $C$ (homogeneous to squared frequencies). Under the action of this transformation the
Hamiltonian can be cast in the form of the sum of two decoupled oscillators, that is,
\begin{equation}
H = \frac{1}{2M} \left( p_+^{\ 2}  + p_-^{\ 2} \right) + \frac{1}{2}M\omega_+^{2} {r}_+^{\ 2} +
\frac{1}{2}M\omega_-^{2} {r}_-^{\ 2},
\end{equation}
and it can be easily shown that
\begin{equation}\label{freqs}
\omega_\pm = \frac{1}{\sqrt{2}} \left[\left( \omega_A^{\ 2} + \omega_B^{\ 2} \right) \pm \left\{ {(\omega_A^{\ 2} -
\omega_B^{\ 2})}^2
+  \frac{4\beta^{\ 2}}{M^2 R^6} \right\}^{1/2} \right]^{1/2}.
\end{equation}
By considering ${r}_A$, ${r}_B$, ${p}_A$ and ${p}_{B}$ as observables subject to standard commutation relations, the energy values of the system are obviously given by
\begin{equation}\label{energie}
E = \hbar\omega_{+} \left( n_{+} + \frac{1}{2} \right) + \hbar\omega_{-} \left( n_{-} + \frac{1}{2} \right)
\end{equation}
where $n_{+},n_{-} \in \mathbb{N}$, that is, are integers. Now, let us consider two opposite physical situations
depending on the relative values of the frequencies $\omega_A$ and $\omega_B$ of the oscillators.

\begin{enumerate}
 \item Consider $\omega_A \gg \omega_B$ (or, equivalently, $\omega_A \ll \omega_B$), we have
\begin{equation}
\omega_\pm = \frac{1}{\sqrt{2}} \left[\left( \omega_A^{\ 2} + \omega_B^{\ 2} \right) \pm \left(\omega_A^{\ 2} -
\omega_B^{\ 2} \right) \left\{  1 +  \frac{4\beta^{\ 2}}{(\omega_A^{\ 2} - \omega_B^{\ 2})^2 M^2 R^6} \right\}^{1/2} \right]^{1/2},
\end{equation}
and the denominator of the last term is large enough to give at the lowest order expansion
\begin{equation}\nonumber
\begin{array}{lll}
\omega_\pm & = & \frac{1}{\sqrt{2}} \left[\left( \omega_A^{\ 2} + \omega_B^{\ 2} \right) \pm {(\omega_A^{\ 2} -
\omega_B^{\ 2})}    \pm  \frac{2\beta^{\ 2}}{(\omega_A^{\ 2} - \omega_B^{\ 2}) M^2 R^6}  +... \right]^{1/2} \\
&&\\
& = & \frac{1}{\sqrt{2}} \left[2 \omega_{A,B}^{\ 2} \pm  \frac{2 \beta^{\ 2}}{(\omega_A^{\ 2} - \omega_B^{\ 2})
 M^2 R^6} +...  \right]^{1/2}  = \omega_{A,B}   \pm
\frac{\beta^{\ 2}}{2\omega_{A,B} (\omega_A^{\ 2} - \omega_B^{\ 2})M^2 R^6}  +...
\end{array}
\end{equation}
where $\omega_{A,B}$ stands for $\omega_{A}$ in the computation of $\omega_+$ and $\omega_{B}$ in the computation of  $\omega_-$.
By substituting this expression  in Eq.(\ref{energie}) we get
\begin{equation}
\begin{array}{rcl}
E &=& \hbar \omega_A  \left( n_{+} + \frac{1}{2} \right) + \hbar\omega_{B} \left( n_{-} + \frac{1}{2} \right)+
\\
\\
&&\ \ \ \ \ \ \frac{\hbar\beta^{\ 2}}{2(\omega_A^{\ 2} - \omega_B^{\ 2})M^2 R^6} \left\{ \frac{1}{\omega_A}
\left( n_{+} + \frac{1}{2} \right) - \frac{1}{\omega_B}  \left( n_{-} + \frac{1}{2} \right) \right\} + ...
\end{array}
\end{equation}
The first two terms correspond to the unperturbed energies of the oscillators $A$ and $B$ considered as isolated
($R\rightarrow \infty$) while the last term provides the lowest order correction to the unperturbed energy of the
system and due to the interaction, this interaction potential energy is proportional to $R^{-6}$. Note that this is functionally the same as the London - Van der Waals interaction but of a remarkably different physical origin
(real oscillations instead of quantum fluctuations).

\item To the contrary, at resonance, that is, $\omega_A \simeq \omega_B = \omega$, the eigen-frequencies (\ref{freqs})
are simply given by
\begin{equation}\label{resonance}
\omega_\pm =  \omega\sqrt{ 1  \pm  \frac{\beta}{M \omega^{\ 2} R^3}} .
\end{equation}
At long distances (imposed by the reality condition for $\omega_\pm$ in this equation) we can develop $\omega_\pm$
near $\omega$ and replace such a development into Eq.(\ref{energie}) to obtain
\begin{equation}\label{frohlich}
E = \hbar\omega ( n_{+}+n_{-}+ 1) + \frac{\hbar \beta}{ 2 M \omega R^3} \left ( n_{+} - n_{-} \right) -
\frac{\hbar\beta^2}{ 8  M^2 \omega^3 R^6} \left( n_{+} - n_{-} + 1 \right) + ...
\end{equation}
The first order correction to the energy of the system corresponds to the interaction energy between the two
oscillators at resonance and is proportional to $R^{-3}$.
If both oscillators are in their ground states, i.e. $n_{+} = n_{-} =0$, the first contribution to
the interaction energy in Eq.(\ref{frohlich}) vanishes as well as the force given above. The first non vanishing term
is again proportional to $R^{-6}$.
But if the lowest of these modes ($\omega_{-}$) gets more excited than the other ($\omega_{+}$) then the consequence is the activation of an {\it attractive} long-range frequency-selective force. A repulsive force could also be activated in case $n_{+} > n_{-}$.

In the context of Fr\"ohlich's theory \cite{pokornybook,frohlich1,ieee,frohlich5} the above described mechanism of resonant interaction between oscillating
dipoles was surmised to have a great potential relevance for fundamental biological processes at the molecular level. Fr\"ohlich proposed a model describing the coupling between the elastic vibrations of macromolecules and the resulting time variations of their dipole moment; the model predicts that one or a few Fourier modes of the dipole field oscillation
should be strongly (coherently) excited provided that the energy supply rate exerted on the macromolecule by its environment exceeds a threshold value. This energy supply is assumed to depend on the biological activity of the environment (metabolic energy). The strongly excited mode of oscillation of
the molecular dipole moment  should be due a collective oscillation either of the entire molecule or of
a subgroup of its atoms.
The consequence of such collective oscillations would be to activate selective long-range recognition and attraction between cognate macromolecular partners via the above described mechanism of resonant interaction.
Experimental evidence of the existence of collective excitations in macromolecules of biological relevance is available for polynucleotides (DNA and RNA) \cite{dna} and for proteins \cite{proteins} in the Raman and far infrared (TeraHertz) spectroscopic domains.

\end{enumerate}


\begin{thebibliography}{99}
\bibitem{riggs} A.D. Riggs, S. Bourgeois, and M. Cohn, J. Mol. Biol. {\bf 53}, 401 (1970).
\bibitem{barkley} M.D. Barkley, Biochemistry {\bf 20}, 3833 (1981).
\bibitem{hippel1} O.G. Berg, R.B. Winter, and P.H. von Hippel, Biochemistry {\bf 20}, 6929 (1981).
\bibitem{hippel2} O.G. Berg, and P.H. von Hippel, Ann. Rev. Biophys. Biophys. Chem {\bf 14}, 131 (1985).
\bibitem{hippel3} O.G. Berg, and P.H. von Hippel, J. Biol. Chem. {\bf 264}, 675 (1989).
\bibitem{cherstvy} A.G. Cherstvy, A.B. Kolomeiski, and A.A. Kornyshev, J. Phys. Chem. B{\bf 112}, 4741 (2008).
\bibitem{cifra} See the review paper :  M. Cifra, J.Z. Fields and A. Farhadi, Prog. Biophys. Mol. Biol. {\bf 105}, 223 (2011).
\bibitem{pokornybook} J. Pokorn\'y and Tsu-Ming Wu, {\it Biophysical Aspects of Coherence and Biological Order}, (Springer, Berlin, 1998).
\bibitem{watson} B. Alberts, D. Bray, J. Lewis, M. Raff, K. Roberts and J. D. Watson, {\it Molecular Biology of the Cell}, (Garland, New York, 1983).
\bibitem{craig} D.P. Craig and T. Thirunamachandran, {\it Molecular Quantum Electrodynamics}, (Academic, London, 1984).
\bibitem{roberto} G. Compagno, R. Passante, and F. Persico, {\it Atom-Field Interactions and Dressed Atoms},
(Cambridge University Press, Cambridge, 1995).
\bibitem{stephen} M. J. Stephen, J. Chem. Phys. {\bf 40}, 669 (1964).
\bibitem{mcLachlan} A.D. McLachlan, Molecular Phys. {\bf 8}, 409 (1964).
\bibitem{jordan} P. Jordan, Phys. Z. {\bf 39}, 711 (1938); P. Jordan, Z. Phys. {\bf 113}, 431 (1939).
\bibitem{pauling} L. Pauling, Science {\bf 92}, 77 (1940).
\bibitem{pauling2} L. Pauling, Nature {\bf 248}, 769 (1974).
\bibitem{Jehle} H. Jehle, Proc. Natl. Acad. Sci. U.S.A. {\bf 50}, 516 (1963).
\bibitem{frohlich1} H. Fröhlich, Int. J. Quantum Chem. {\bf 2}, 641 (1968).
\bibitem{nota1} A typical example of the existence of non thermal behaviors in living matter at the microscopic level is provided by basic energy conversion mechanisms.
According to the estimates provided by electrochemistry (see, for example, J. Bockris, and S. Khan, {\it Surface Electrochemistry} (Plenum Press, New York,
1993), Chapter 7), the efficiency of energy production in mammals and humans is very high: about $50\%$.
On the other hand, higher living organisms are at a temperature $T$ slightly above $300\text{K}$ with an excursion $\Delta T$ of a few degrees, whence - according to the second law of thermodynamics - the thermodynamic (equilibrium) efficiency $\Delta T/T$ should be about $1\%$, much lower indeed. This is a quantitative example of why fundamental
processes in living matter at the molecular level must stem from a strongly correlated and coherent dynamics.
\bibitem{frohlich2} H. Fröhlich, Phys. Lett. A{\bf 39}, 153 (1972).
\bibitem{frohlich3} H. Fröhlich, Proc. Natl. Acad. Sci. U.S.A. {\bf 72}, 4211 (1975).
\bibitem{frohlich4} H. Fröhlich, Advances in Electronics and Electron Physics {\bf 53}, 85-152 (1980).
\bibitem{rowlands} S. Rowlands, L.S. Sewchand, R.E. Lovlin, J.S. Beck and E.G. Enns, Phys. Lett. A{\bf 82}, 436 (1981); S. Rowlands, L.S. Sewchand, and E.G. Enns, Phys. Lett. A{\bf 87}, 256 (1982).
\bibitem{tuszynski} R. Paul, R. Chatterjee, J.A. Tuszynski, and O.G. Fritz, J. Theor. Biol. {\bf 104}, 169 (1983).
\bibitem{reimers} J. Reimers, L. McKemmish, A. Mark, R. McKenzie, and N. Hush, Proc. Natl. Acad. Sci. U.S.A. {\bf 106}, 4219 (2009).
\bibitem{proteins} See for instance: P.C. Painter, L.E. Mosher, and C. Rhoads, Biopolymers {\bf 21}, 1469 (1982); K.-C. Chou, Biophys. J. {\bf 48}, 289 (1985); A. Xie, A.F.G. van der Meer, and R.H. Austin, Phys. Rev. Lett. {\bf 88}, 018102 (2002), and references quoted in these papers.
\bibitem{dna} P.C. Painter, L.E. Mosher, and C. Rhoads, Biopolymers {\bf 20}, 243 (1981); H. Urabe, and Y. Tominaga, Biopolymers {\bf 21}, 2477 (1982); K.-C. Chou, Biochem. J. {\bf 221}, 27 (1984); J.W. Powell, et al., Phys. Rev. A{\bf35}, 3929 (1987); B.M. Fisher, M. Walther, and P.U. Jepsen, Phys. Med. Biol. {\bf 47}, 3807 (2002), and references quoted in these papers.
\bibitem{Pettinibook} M. Pettini, {\it Geometry and Topology in Hamiltonian Dynamics and Statistical Mechanics}, IAM Series n. 33, (Springer, New York, 2007).
\bibitem{elena1} E. Floriani, R. Lima, Chaos {\bf 9}, 715 (1999).
\bibitem{elena2} E. Floriani, D. Volchenkov, R. Lima, J. Phys. A: Math. Gen. {\bf 36}, 4771 (2003).
\bibitem{Gardiner} C.W. Gardiner, {\it Handbook of Stochastic Methods}, (Springer-Verlag, Berlin, 1985);
N.G. Van Kampen, {\it Stochastic Processes in Physics and Chemistry}, (North-Holland, Amsterdam, 1981).
\bibitem{ellison} W. J. Ellison, J. Phys. Chem. Ref. Data {\bf 36}, No. 1, 1 (2007).
\bibitem{shure} L. Shure, {\it Two-dimensional integration over a general domain}
(\url{http://blogs.mathworks.com/loren/2006/04/26/two-dimensional-integration-over-a-general-domain}), MathWorks (2006).
\bibitem{debye} P. Debye, Trans. Electrochem. Soc. {\bf 82}, 265 (1942).
\bibitem{noyes} R.M. Noyes, Prog. React. Kinet. {\bf 1}, 129 (1961).
\bibitem{nota2} Further analysis made in presence of other various potential revealed an
asymptotic $x^{n+2}$ dependence for $\tau$ at small $x$-values when $U(x) \propto - x^{-n}$.
\bibitem{maiti} S. Maiti, U. Haupts, and W.W. Webb, Proc. Natl. Acad. Sci. U.S.A. {\bf 94}, 11753 (1997).
\bibitem{hom} E.F. Hom, A.S. Verkman,  Biophys. J. {\bf 83}, 533 (2002).
\bibitem{schwille} K. Bacia, P. Schwille, Methods {\bf 29}, 74 (2003).
\bibitem{ww} B. Mugnier, B. Nal, C. Verthuy, C. Boyer, D. Lam, L. Chasson, V. Nieoullon, G. Chazal, X-J. Guo, H-T. He, D. Rueff-Juy, A. Alcover, P. Ferrier, PLoS ONE {\bf 3}, e3467 (2008); A. Pekowska, T. Benoukraf, P. Ferrier, S. Spicuglia, Genome Res. {\bf 20}, 1493 (2010).
\bibitem{xx} F. Koch, R. Fenouil, M. Gut, P. Cauchy, T.K. Albert, J. Zacarias-Cabeza, S. Spicuglia, A.L. de la Chapelle, M. Heidemann, C. Hintermair, D. Eick, I. Gut, P. Ferrier, J.C. Andrau, Nature Struct. Mol. Biol. {\bf 18}, 956 (2011); A. Pekowska, T. Benoukraf, J. Zacarias-Cabeza, M. Belhocine, F. Koch, H. Holota, J. Imbert, J.C. Andrau, P. Ferrier, S. Spicuglia, EMBO J. (2011) Aug 16. doi: 10.1038/emboj.2011.295.
\bibitem{yy} A. Bancaud, et al., EMBO J. {\bf 28}, 3785 (2009).
\bibitem{zz} A. Chaudhuri, et al., Proc. Natl. Acad. Sci. USA {\bf 108}, 14825 (2011).
\bibitem{seshadri} The mean first passage time for Wiener-Einstein processes to attain a given absolute displacement
 is found to be independent of the dimensionality of the process in: V. Seshadri and K. Lindenberg, J. Stat. Phys. {\bf 22}, 69 (1980).
\bibitem{israel} J. N. Israelachvili, Quart. Rev. Biophys. {\bf 6}, 341 (1974).
\bibitem{margenau} H. Margenau, Rev. Mod. Phys. {\bf 11}, 1 (1939).
\bibitem{pretopettini} A more refined treatment of this problem is given in: J. Preto and M. Pettini, {\it Long range resonant interactions in biological systems}, (2011) preprint.
\bibitem{jackson} J.D. Jackson, {\it Classical Electrodynamics},(John Wiley \& Sons, New York, 1975).
\bibitem{ieee} H. Fr\"ohlich, IEEE Trans. Microwave Theor. \& Techn. {\bf 26}, 613 (1978).
\bibitem{frohlich5} H. Fröhlich, Rivista Nuovo Cimento {\bf 7}, 399 (1977).
\end{thebibliography}
\end{document}